# Game of the Cursed Prince based on Android

Risky Armansyah
Trisakti University
Indonesia

Dian Pratiwi
Trisakti University
Indonesia

## ABSTRACT
Nowadays Games become an entertainment alternative for various circles, industry and game development business is also a profitable industry. In Indonesia the amount of game consumption is very high, especially the console game type RPG (Role Playing Game). The task of this research is developing game software using Unity3d to create an Android-based RPG game app. The story is packed with RPG genres so the player can feel the main role of the story's imagination. The game to be built is a game titled 'The Cursed Prince'. Users will get the sensation of royal adventure. Multiplayer game system, graphics in 3D game, The main character in this game is Prince, enemies in this game are wizards and monsters, Game is not limited time to complete. And the game can be saved, so it can be reopened. The game of 'The Cursed Prince' can be part of Indonesian Industry Gaming development.

## 1. INTRODUCTION
The development of game technology has now entered the industrial era where, the game has become an alternative entertainment for various circles, from young to adults. Industry and game development business is also a profitable industry development companies in America, Japan, and Europe can reach high profit. Consumption of games in Indonesia is currently increasing especially the game console, many overseas game companies that sell the game to be played in Indonesia. One type of game in the market is Role Playing Game (RPG).

Role Playing Game (RPG) is a game with complex story element and the role makes a user feel like being a character in the game. [1] states, games based on the RPG genre take place in a set of realistic worlds in a particular era (for example, in medieval times, at present or in the future) or in an imaginary world that is not close to reality. Players play the characters represented by avatars (most often shaped humans) and solve various quests throughout the game.

RPG gives the player a quest to find a particular item and use it properly or combine it to solve a particular problem and require the player to choose the correct answer from a certain number of questions for answers. Quests are assigned by non-player characters (NPCs) that can not be controlled by players and interact with through dialogs. In the RPG game model, each player can collect objects or anything that contains the requirements to complete the quest during game play. In addition, game objects that represent learning objects have been defined locations that can be NPCs or other objects.

Unity3d is RPG editing engine where RPG 2D and 3D can be self-contained programs that can be played instantly without help of this program or other programs. Games displayed in this final project based on the story of the author's own imagination. The story of the imagination in recent decades is less well known and lacks creativity by the younger generation. Based on that, the imagination story taken for this game is the exiled Prince of Cursed. The story begins with the Prince being cursed by a witch, this story is an improvised Indonesian folklore. The Prince is cursed to be a monster and given a certain time and a certain way to restore the original form.

This game almost the same with Zenonia App, which is a God in throw to earth to address the earth from the attack of evil monsters. While the story of the Cursed Prince is a battle between the Prince, the Monster and the Witch. In this game there are some chapter or part to be passed by the Cursed Prince. Where the Prince finally wages war against the monsters in advance to gain experience (EXP) and get armor from physical attacks and magic attacks. Monsters in this game have different levels of level each chapter or part against monsters with different levels or higher monsters. After fighting the monsters at the end of the chapter the Prince fights and defeats the Witch to restore the Witch's curse for Pangeraan back to its original human form. The purpose of this study is that the story of the imagination can be known by the public, and the community can be creative in the story. The story is packed with the RPG genre so the player can feel the main role of the imaginary story.

## 2. LITERATURE
According to [1] RPG game is a game where the players play a role to follow a story. The players have distinctive characteristics of each character in the game. RPG games such as a novel or a movie where this is the main attraction because it makes the players imagine as characters in the RPG game. RPG usually leads to social collaboration rather than a competition. Generally in RPG, the players are joined in one group. An RPG game has certain characteristics such as, the player must kill several monsters for his character to be strong, the player can input the name of the character being played, determine the rules of his character battle and his performances in the imaginary world to be used (history, geography, kings name, and others) [1].

Researchers use unity as a game engine to create RPG games. The process of assisted tools and Asset that is available in a program. Unity Technologies was built in 2004 by David Helgason, Nicholas Francis, and Joachim Ante in Copenhagen, Denmark. Unity Technologies, Unity Technologies is the best game engine for developers.

Research using waterfall model in traditional software development process, waterfall commonly used to simplify the work of making software. This is a sequential model, so completion of a set of activities leads to the commencement of subsequent activities [10]. This is called waterfall because the process flows "systematically from one stage to another in the downward mode". Establish a framework for software development. Several variants of the model exist, each different label using for each stage.

The console that can play this game is Android an operating system for linux-based mobile devices that includes operating systems, middleware and applications. Android provides an open platform for developers to create their apps.

Unified Modeling Language (UML) is a language for specification, visualization, development and documentation of software systems. In UML design, the system is defined as a set of objects that have attributes and methods. Attributes are the variables attached to objects and methods are functions that can be done by the object. The object class can not stand alone, in its use, an embodiment (instantiation) of the object is performed.

## 3. METHODOLOGY

The game to be built is a RPG game titled 'The Cursed Prince". This a RPG-based adventure game is using android as a console, in the game there has a moral message of life. Users will get the sensation of adventure through this android based game. Here is an analysis of existing systems in this game, among others: (1) Game genre RPG (Role Playing Game). (2) Single player game system. (3) Graph in 2D game. (4) The main character in this game is Prince. (5) The enemies in this game are wizards and monsters. (7) The game is not limited by time to complete it. (8) The game can be saved, so it can be reopened.

Author collecting data and information on the issues discussed, the authors read and study the results of emerging technologies such as the internet, e-books, films, folklore, and others related to the study of the Cursed Prince's story for reference. System development method used in completing this Final Project using Waterfall model consisting of: (a) Stages of planning conditions at this stage high-level users decide what functions should be featured by the game using the Unity3d. (b) Stages of user design, at this stage done the design process and interface design of the game using Unity3d then convert to Android. (c) Construction Phase, at this stage coding of designs has been defined. (d) Phase of Implementation, at this stage testing and testing analysis of games created using Unity3d

This game is intended for everyone from elementary school children who can read to adults, User Age 7 years - 30 years, has ability can use mobile phone with android operating system, at least user can read and not illiterate. This adventure game app combines three components: the fictional story of Indonesian culture, the fictional story of western culture, and also its own creativity story entitled the Cursed Prince

The functional requirements analysis describes the process of activities to be applied in an application and explains the need for the application to run properly. The software needed to build the RPG Game of The Cursed Prince App is as follows: (1) Windows operating system 8. (2) Unity3d, (3) Audacity (sound), (4) Android SDK (Android Studio).

When building this RPG Game of The Cursed Prince App, the hardware used is as follows: (1) Smartphone with Android operating system, (2) 4GB DDR4 Memory RAM, (3) NVIDIA 940MX 2GB VGA-RAM, (4) Intel Core i5 -6200U, (5) Data Cable and (6) Internet Modem (wired or wireless).

Use case Diagram is a construction to describe the relationships that occur between actors with the activities contained in the application. Target use case modeling among others is to define the functional and operational needs of the system by defining the scenario of system usage to be built. From the results of the analysis of existing applications then use case diagram for the Cursed Prince game application can be viewed in the following figure:

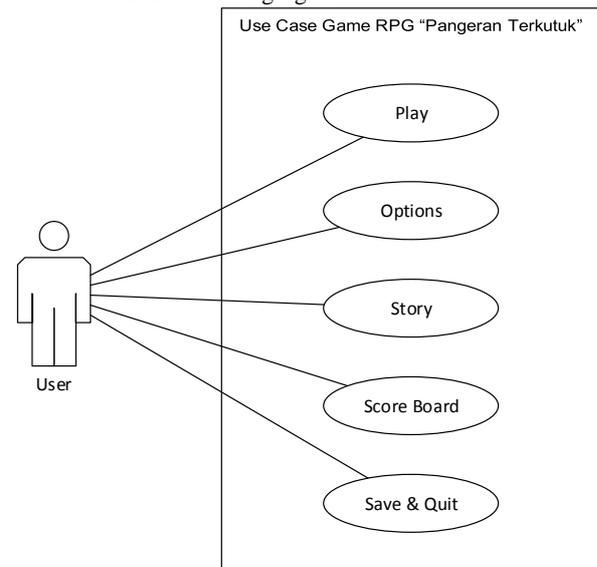

**Fig 1. Use Case Diagram Game The Cursed Prince**

## 4. RESULT AND DISCUSSION

### 4.1 Storyline

Tittle: Pangeran Terkutuk (The Cursed Prince)

A story in a prosperous, rich and happy kingdom has a King, a Queen, a Prince, a concubine, a Treasure, and so on. They are surrounded with a luxurious and happy life. But they do not know what threats will befall them. One day a magician came along. With a gripping atmosphere, magicians enter the palace to make the kingdom destroyed. Inside the king's palace, the Queen, the Prince, and the king's aides finally meet the Witch. The wizard also tells that at one time this kingdom will make the Sorcerer miserable and if want to return as before they must defeat the Sorcerer. Witches also make the kingdom becomes tense and miserable. The Witch's target changed after making the kingdom destroyed, the Prince. Because the Prince is a moment to be that will make him become destroyed. The witch makes the Prince faint and cursed the prince into an ugly monster.

At that moment, the King and Queen see the Prince cursed by the Witch and ask for forgiveness. The King negotiates with witches so that the kingdom and Prince are not destroyed anymore. But the wizard would not listen and remain at his stance. The cursed prince was taken by the guard to be thrown into the forest. After being dumped a few days later the Prince wakes up and is shocked at the shape of himself. Prince is angry, sad, and aloof. After walking the path in the woods, he found a secret box. Prince curious and open the secret box. The Prince opened the box and contained weapons. Prince thought to take revenge against the Witch.

Sometime later the Prince heads to the palace to defeat the Witch. Before defeating the Prince Wizard has been confronted by a monster of monsters in order to protect the Witch from destruction. From the forest to the royal gate monsters guard the witches from the terror of the Cursed Prince. The prince is able to defeat the monster monsters with the help of weapons that Prince finds in the forest. The prince prepares to defeat the Witch.

After defeating the monsters and carrying the necessary tools to defeat the Witch, the Prince returns to the palace to judge the witch in his kingdom. The prince fought dead with the Witch. And in the end the Prince wins against the Witch. Prince is very tired and almost dead. After the victory, the Prince finally sees himself back to normal and the kingdom returns to normal. The King, Queen, Consort, Army, etc. finally get out of their hideout. They also greeted with joy after the Wizard's defeat.

## 4.2 Character Design

In the game Prince of the Cursed there are some characters that portray their roles, namely:

1. Prince

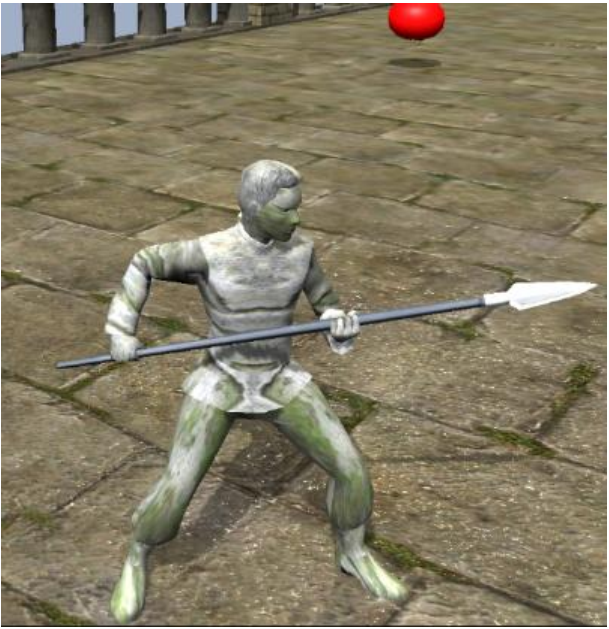

**Fig 2. The Cursed Prince**

Prince (Fig 2) is a good lead character in the game of the Cursed Prince. Prince lives in the kingdom with his family in peace and happiness. But because the kingdom of the coming witch who later changed himself, finally the prince must be out of work to the forest. In order to return to the kingdom the prince must defeat the witch.

2. The Witch

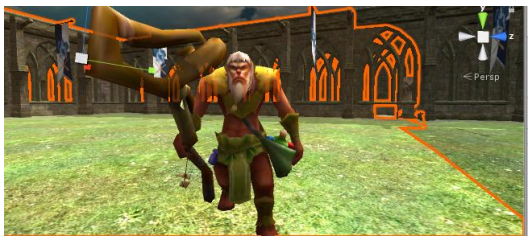

**Fig 3. Witch.**

Wizard (figure 3) is the second evil character in the game of the Cursed Prince. Witches one day come to the Prince in the kingdom and conjure him from human beings into monsters. Witches ordered monsters to block the prince's intention to kill the witch. The witch is eventually defeated by the prince, his magic spell is lost and the prince returns to his original form.

3. Monster

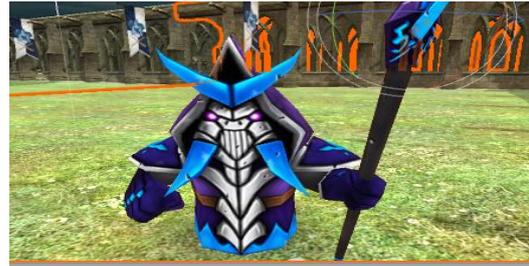

**Fig 4. Monster**

The monster (figure 4) is a subordinate of a wizard who is commanded to block the prince's intention to kill a witch. But in the end the monster was defeated by the prince.

4. King

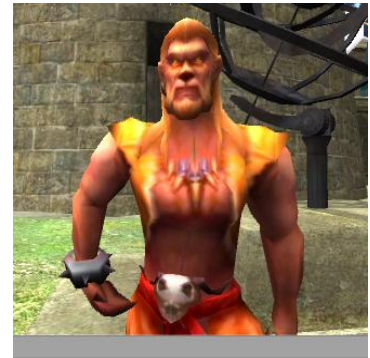

**Fig 5. King**

The king (figure 5) is the figure who leads the kingdom and father of the damned Prince. The king embarked on the negation of the Witch who, according to the King, harmed himself and the kingdom.

## 4.3 Hardware dan Software

Overall this can be implemented using the specs as written below:

To implement the features on the system that has been designed, then the hardware specifications (hardware) used are as follows: (a) Intel Core i3 minimal processor, (b) Minimum speed 2.3GHz, (c) Memory RAM (Random Access Memory) of at least 4096Mb, (d) Hard disk drive minimum space 8Gb, (e) 1GB of VGA RAM and (f) Samsung S7 Edge

In addition to hardware, in order to be able to run properly it must be supported by the software (software). The software used is as follows (a) Windows 10 64-bit, (b) Unity 5.5.0 in game creation and (c) Android OS 7.0

## 4.4 Game 'The Cursed Prince'

There are menu options such as play games, continue, multiplayer, about, and exit.

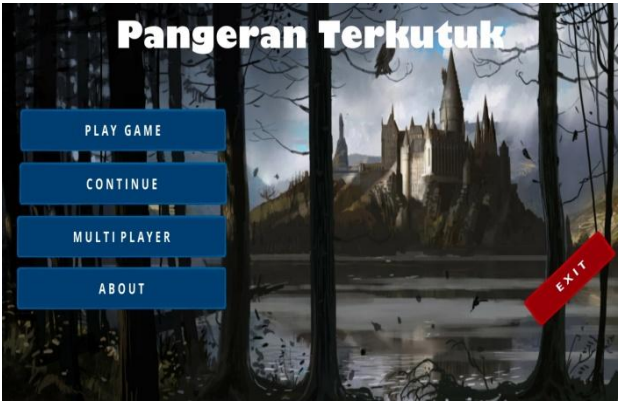

**Fig 6. Main Page.**

Start a game that already has a storyline in game and continue is a resume games that have been saved automatically.

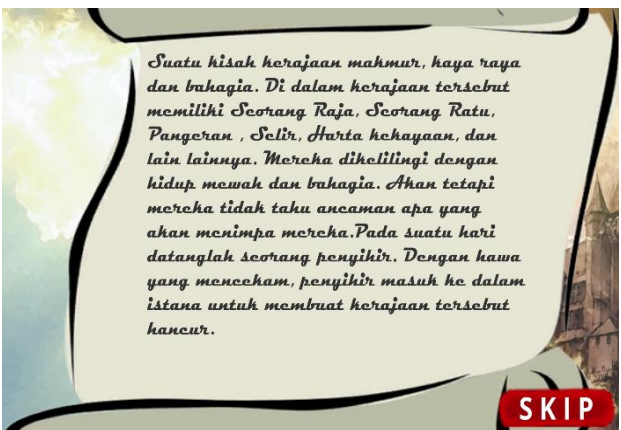

**Fig.7 Narration in game.**

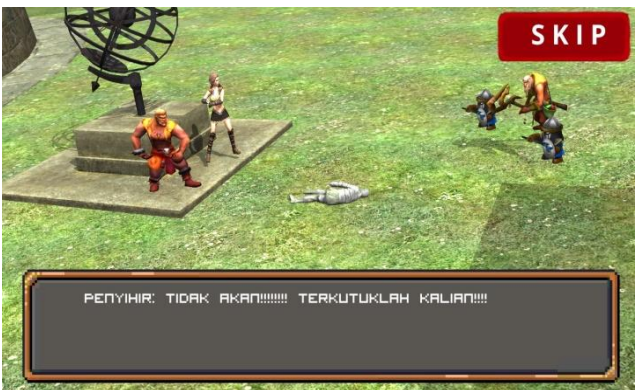

**Fig 8. Dialog in game.**

Scene selection of weapons of the main character, after selecting the armed character in the end the RPG game has been played with different scene level different.

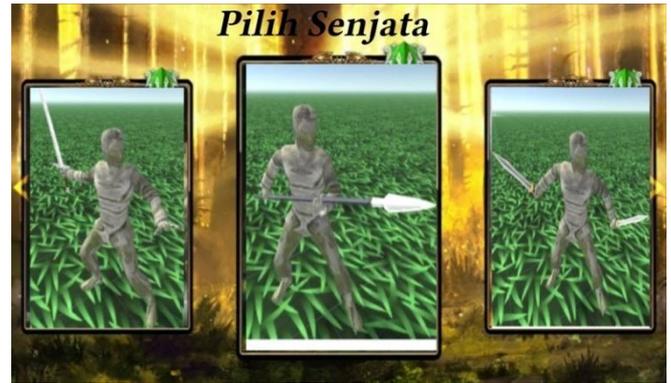

**Fig 9. Character Selection.**

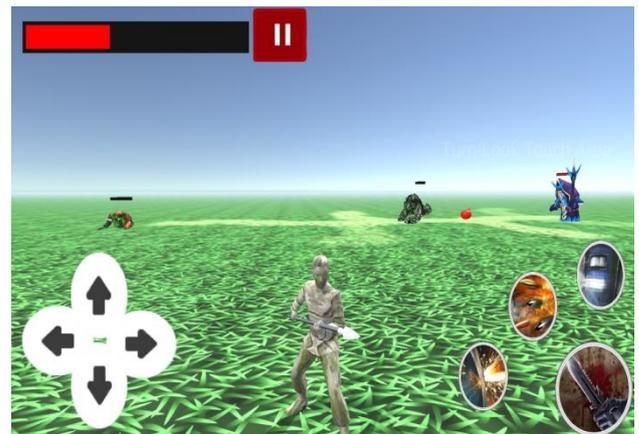

**Fig 10. Game Scene Level 1**

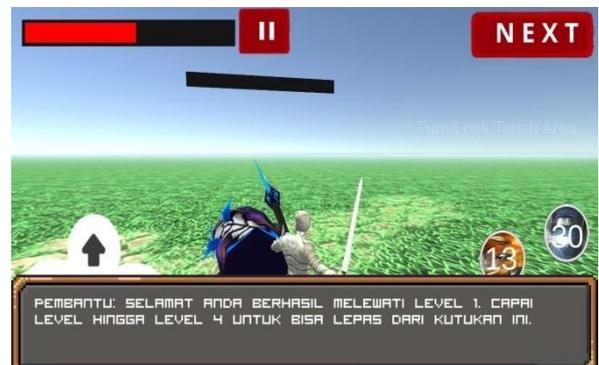

**Fig 11. Game Scene Level 1 Information.**

Each completed game scene level players will get the following information.

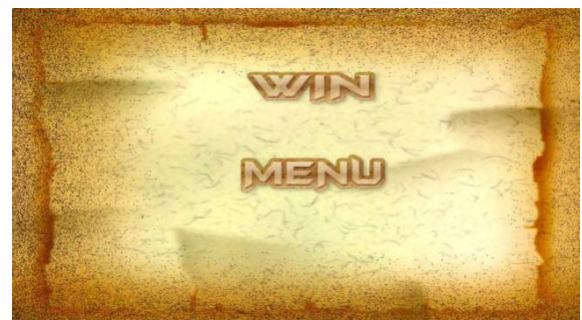

**Fig 12. Next Chapter Information.**

The game is over and will return to the main page.

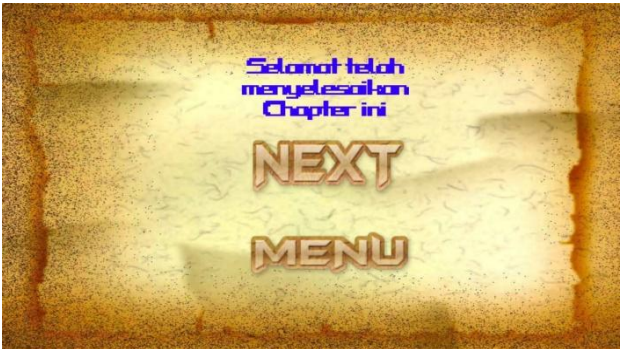

**Fig 13. Win Scene.**

The lost game will get the information and continue the game will repeat from the game scene level that has been the last player reach or return to the main page

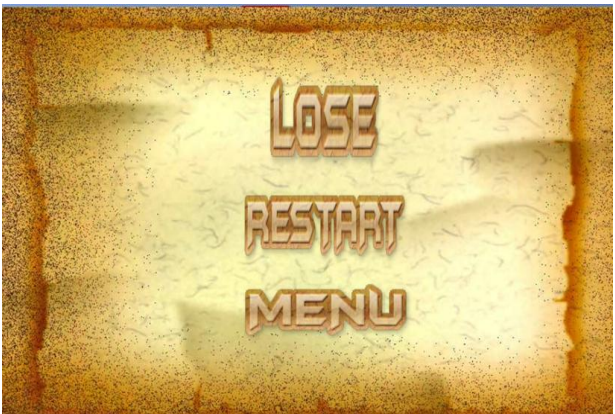

**Fig 14. Lose Scene.**

In this Menu the game will be played by not following the story line and just fighting against the monsters available in the game.

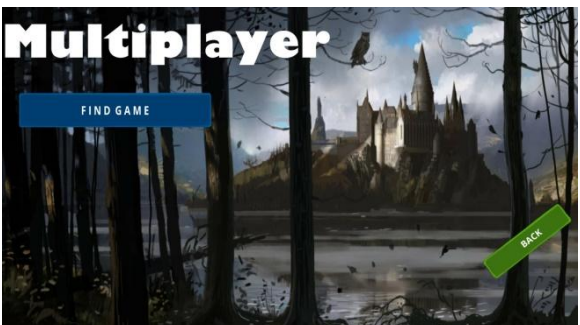

**Fig 15. Multiplayer Menu.**

A menu that contains the End Task Creator.

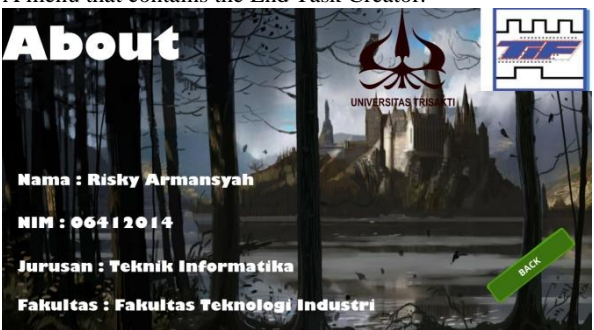

**Fig 16. Menu About**

Exit game menu for the quit game and back to home of smartphone

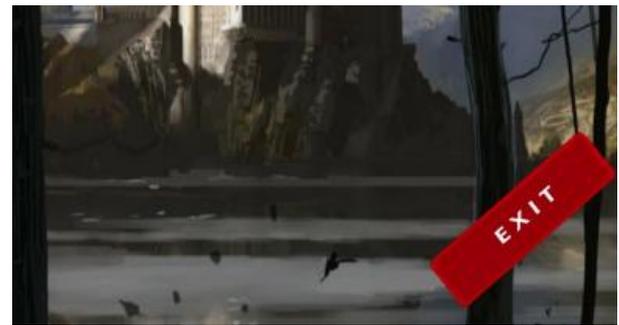

**Fig 17. Exit Menu.**

## 5. CONCLUSION
The conclusions that can be taken in the work of this Cursed RPG Game app are (1) Creating a Cursed Prince RPG game using Unity3d and designed with a creative fictional story that has been prepared. Inside this RPG game has meaning and purpose for the players. (2) Create multiplayer game must have own server and database. Players will connect with each other, and defeat monsters simultaneously that have been put up in the game. After making the RPG game, then the suggestion given is expected one time this application can be developed in terms of systems, programs, and features to make it easier in use. This RPG game can all can enable less features such as multiplayer. (3) Helps develop the player's overall personality-how the player will act on a particular situation, how the player decides case examples, and solves problems.

## 6. ACKNOWLEDGMENTS
Thank you to my family for helping me materially.